\documentclass[lettersize,journal]{IEEEtran}
\usepackage{times}
\usepackage{soul}
\usepackage{url}
\usepackage[utf8]{inputenc}
\usepackage[small]{caption}
\usepackage{graphicx}
\usepackage{amsmath}
\usepackage{amsthm}
\usepackage{booktabs}
\usepackage{algorithm}
\usepackage{algorithmic}
\usepackage[switch]{lineno}
\usepackage{latexsym}

\usepackage{amsmath,amsfonts}
\usepackage{array}
\usepackage[caption=false,font=normalsize,labelfont=sf,textfont=sf]{subfig}
\usepackage{textcomp}
\usepackage{stfloats}
\usepackage{url}
\usepackage{verbatim}

\usepackage{bm}
\usepackage{color}
\usepackage{multirow}
\usepackage{caption}
\usepackage{amstext}
\usepackage{setspace}

\begin{document}

\title{Multimodal Matching-aware Co-attention Networks with Mutual Knowledge Distillation for Fake News Detection}

\author{Linmei~Hu,~Ziwang~Zhao,
        ~Weijian~Qi,~Xuemeng~Song,\emph{Senior Member, IEEE}\\
        and~Liqiang~Nie, \emph{Senior Member, IEEE}

\thanks{This paper was produced by the IEEE Publication Technology Group. They are in Piscataway, NJ.}
\thanks{Manuscript received April 19, 2021; revised August 16, 2021.}}

\markboth{Journal of \LaTeX\ Class Files,~Vol.~14, No.~8, August~2021}%
{Shell \MakeLowercase{\textit{et al.}}: A Sample Article Using IEEEtran.cls for IEEE Journals}


\maketitle

\begin{abstract}
Fake news often involves multimedia information such as text and image to mislead readers, proliferating and expanding its influence. 
Most existing fake news detection methods apply the co-attention mechanism to fuse multimodal features while ignoring the consistency of image and text in co-attention.  
In this paper, we propose multimodal matching-aware co-attention networks with mutual knowledge distillation for improving fake news detection.
Specifically, we design an image-text matching-aware co-attention mechanism which captures the alignment of image and text for better multimodal fusion. The image-text matching representation can be obtained via a vision-language pre-trained model. Additionally, based on the designed image-text matching-aware co-attention mechanism, we propose to build two co-attention networks respectively centered on text and image for mutual knowledge distillation to improve fake news detection.
Extensive experiments on three benchmark datasets demonstrate that our proposed model achieves state-of-the-art performance on multimodal fake news detection.
\end{abstract}


\section{Introduction}
Online social media has become an indispensable platform for people to share and access information in their daily life. 
Due to the loose constraints for user-generated content on social media, there could be plenty of fake news that distorts and fabricates facts, which could mislead readers and even cause a great negative impact on society. Fake news usually uses multimedia information such as text and image to draw users' attention and expand its influence. It has become urgent and important to detect multimodal fake news on social media.

\begin{figure*}[!tbp]
    \centering
    \includegraphics[width=0.9\textwidth]{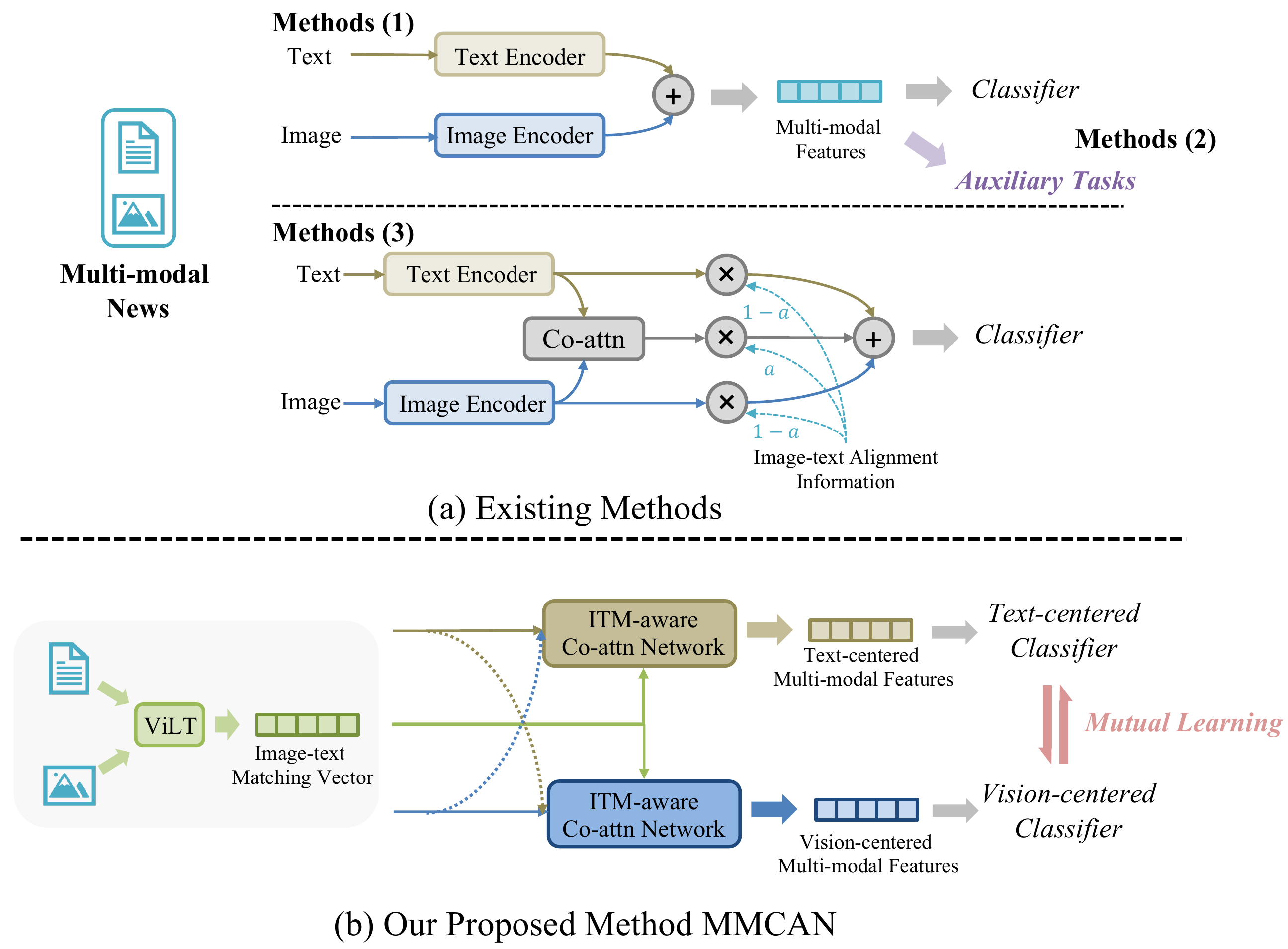}
    \caption{Comparison between (a) existing methods and (b) our proposed method MMCAN.}
    \label{fig:compare}
\end{figure*}

Many efforts have been made on fusing textual and visual features for multimodal fake news detection. 
As shown in Figure \ref{fig:compare}(a), existing studies conduct multimodal fusion via simple concatenation, auxiliary tasks, or co-attention mechanism.
Early studies \cite{DBLP:conf/mm/JinCGZL17,DBLP:conf/bigmm/SinghalS0KS19,DBLP:conf/aaai/SinghalKSS0K20} combine the two modality features by simple concatenation. 
Some studies improve them by introducing auxiliary tasks such as feature reconstruction \cite{DBLP:conf/www/KhattarG0V19} and event discrimination \cite{DBLP:conf/kdd/WangMJYXJSG18} to enhance multimodal feature learning. 
In order to further capture the inter-modality correlations, recent studies adopt co-attention networks for fine-grained modality interactions to detect fake news \cite{DBLP:conf/acl/WuZZWX21,DBLP:conf/mm/0005C0LSMHLGY21,DBLP:conf/ijcai/ZhengZGWZ022}. 
However, they fail to consider the matching degree of the image and text in the co-attention mechanism for multimodal fusion. Fake news could have mismatched images and text, in which case directly applying co-attention between the different modalities could lead to under-performed multimodal features for fake news detection. 

To address the limitation, we propose a novel image-text matching (ITM) aware co-attention mechanism to capture the matching degree of image and text while learning multimodal fusion features. 
Although some existing works also pay attention to the consistency of different modality content in the news by combining the consistency based additional features with multimodal features for fake news detection \cite{DBLP:conf/pakdd/ZhouWZ20,DBLP:journals/ipm/XueWTLSW21,DBLP:conf/www/0003LZSLTS22}, they do not consider the image-text alignment in learning multimodal fusion features that play a critical role in fake news detection. 
Additionally, we propose to conduct mutual learning  between two ITM-aware co-attention networks that are respectively centered on text (text features as query) and image (image features as query), which enables them to learn collaboratively with mutual knowledge distillation for  improving fake news detection.

Overall, in this paper, we propose novel \textbf{M}ultimodal \textbf{M}atching-aware \textbf{C}o-\textbf{A}ttention \textbf{N}etworks with mutual knowledge distillation (MMCAN, shown in Figure \ref{fig:compare}(b)) for improving multimodal fake news detection. Specifically, we first obtain the  ITM representation via a vision-language pre-trained  model such as ViLT \cite{DBLP:conf/icml/KimSK21}. 
Then we design a new ITM-aware co-attention mechanism, which can learn better multimodal fusion features relying on the alignment of image and text in the news. Based on the proposed ITM-aware co-attention, we build two co-attention networks respectively centered on text (text features as query) and image (image features as query) for multimodal fake news detection.
Moreover, mutual learning between two co-attention networks is exploited to enable  knowledge distillation from each other for collaboratively improving fake news detection.
Extensive experiments on three benchmark datasets demonstrate that our model achieves state-of-the-art performance in multimodal fake news detection.
In summary, our main contributions are as follows:

\begin{itemize}

\item We propose novel multimodal matching-aware co-attention networks with mutual knowledge distillation for improving fake news detection. 

\item We design a new ITM-aware co-attention mechanism for learning better multimodal fusion features, guided by the alignment of image and text in the news. Moreover, mutual knowledge distillation of two co-attention networks based on the new co-attention mechanism is also employed to further improve fake news detection.

\item Extensive experiments demonstrate that our model MMCAN achieves state-of-the-art performance on three public benchmark datasets for the multimodal fake news detection task.

\end{itemize}

\section{Related Work}


\subsection{Fake News Detection}
Fake news is the news that is intentionally fabricated and can be verified as fake \cite{DBLP:conf/cikm/RuchanskySL17,shu2017fake}. Existing studies on fake news detection can be divided into unimodal methods and multimodal methods.

\subsubsection{Unimodal Content Based Methods.} The unimodal fake news detection methods can be further divided into two types: textual feature based and visual feature based methods.
In early studies, the textual features are mostly hand-crafted \cite{DBLP:conf/www/CastilloMP11,DBLP:conf/icmi/ChenCR15}, and it is difficult to fully mine the deep semantic information conveyed by the text. To address this problem, many studies use deep learning technologies to learn the textual representation of news towards identifying fake news \cite{DBLP:conf/ijcai/MaGMKJWC16,DBLP:conf/acl/HuYZZTSD020,DBLP:conf/aaai/DunTCHY21,9511228}. For instance, \textit{Liao et al.} \cite{9339883} proposed a graph based method for learning news representations which capture news relations.

The visual feature have been recognized as an important indicator for fake news detection. \textit{Jin et al.} \cite{DBLP:journals/tmm/JinCZZT17} extracted several visual features to characterize image distribution patterns for detecting fake news.
To avoid feature engineering, \textit{Qi et al.} \cite{DBLP:conf/icdm/QiCYGL19} utilized a visual neural network to effectively capture and fuse the characteristics of fake-news images at both physical and semantic levels.



\subsubsection{Multimodal Content Based Methods}
Multimodal fusion features have been shown to play a critical role in fake news detection.  
\textcolor{black}{Early studies \cite{DBLP:conf/mm/JinCGZL17,DBLP:conf/bigmm/SinghalS0KS19,DBLP:conf/aaai/SinghalKSS0K20}} mainly focus on designing more advanced feature extractors for multiple modalities, and then the multimodal fusion is fulfilled simply by the concatenation operation. 
Some studies utilize auxiliary tasks such as feature reconstruction \cite{DBLP:conf/www/KhattarG0V19} and event discrimination \cite{DBLP:conf/kdd/WangMJYXJSG18} to enhance the multimodal feature learning for fake news detection. 
To model the interactions between the two modalities sufficiently, co-attention based methods have been proposed for multimodal fake news detection \cite{DBLP:conf/acl/WuZZWX21,DBLP:conf/sigir/QianWHFX21,DBLP:conf/ijcai/ZhengZGWZ022}.
Concretely, \textit{Qian et al.} \cite{DBLP:conf/sigir/QianWHFX21} considered the hierarchical semantics of text and utilized multiple co-attention layers to perform multimodal interactions. \textit{Zheng et al.} \cite{DBLP:conf/ijcai/ZhengZGWZ022} proposed to fuse textual, visual, and social modal features via the co-attention mechanism. Some works also pay attention to the consistency between the text and image, considering news pieces with mismatched text and image are more likely to be fake than those with matching text and image \cite{DBLP:conf/pakdd/ZhouWZ20,DBLP:journals/ipm/XueWTLSW21,DBLP:journals/tmm/LiSYTYX22,DBLP:conf/www/0003LZSLTS22}. 
For example, \textit{Chen et al.} \cite{DBLP:conf/www/0003LZSLTS22} estimated the cross-modal ambiguity with the Kullback-Leibler (KL) divergence between the unimodal feature distributions and used the ambiguity score to govern the aggregation of unimodal features and multimodal features for fake news detection. \textit{}{Zhou et al.} \cite{DBLP:conf/pakdd/ZhouWZ20} measured the cosine similarity between text and image features for fake news detection, in addition to multimodal feature based fake news detection.

Different from the above works, to learn better multimodal fusion features, we design a new ITM-aware co-attention mechanism capturing the image-text alignment. Moreover, we exploit the mutual learning of two ITM-aware co-attention networks  for  improving fake news detection.

\subsection{Deep Mutual Learning}
Model distillation is an effective and widely used technique to transfer knowledge from a teacher network to a student one. Nevertheless, in practice, there could be no teacher but only students.
Towards this, \textit{Zhang et al.} \cite{DBLP:conf/cvpr/ZhangXHL18} proposed the deep mutual learning, which aims to distill knowledge between peer students by pushing them to learn collaboratively and teach each other. Since then, mutual learning has attracted many researchers’ attention \cite{DBLP:journals/tcsv/WangLSLT22,DBLP:conf/icassp/WeiPQNDL22,DBLP:journals/tcsv/JiangZK21,DBLP:journals/tcsv/ZhangZJWXQ22}.
For example, \textit{Wei et al.} \cite{DBLP:conf/icassp/WeiPQNDL22} introduced distillation loss between textual and visual networks to learn modality correlations for facilitating fake news detection.
\textit{Zhang et al.} \cite{DBLP:journals/tcsv/ZhangZJWXQ22} performed mutual learning between RGB and IR modality branches to improve the intra-modality discrimination in the cross-modal person re-identification task.
Differently, in this work, based on our ITM-aware co-attention mechanism, we build two co-attention networks respectively centered on text and image, and apply mutual learning between them to collaboratively improve fake news detection.

\section{METHODOLOGY}
We propose novel multimodal matching-aware co-attention networks with mutual knowledge distillation (MMCAN) to improve the performance of the fake news detection task. 
In MMCAN, a new multimodal matching-aware co-attention mechanism is designed to capture the matching information of image and text for learning multimodal representations. Based on the new co-attention mechanism, two co-attention networks centered on different modalities are employed to enable mutual knowledge distillation for improving fake news detection.

\begin{figure*}[!t]
    \centering
    \includegraphics[width=0.99\textwidth]{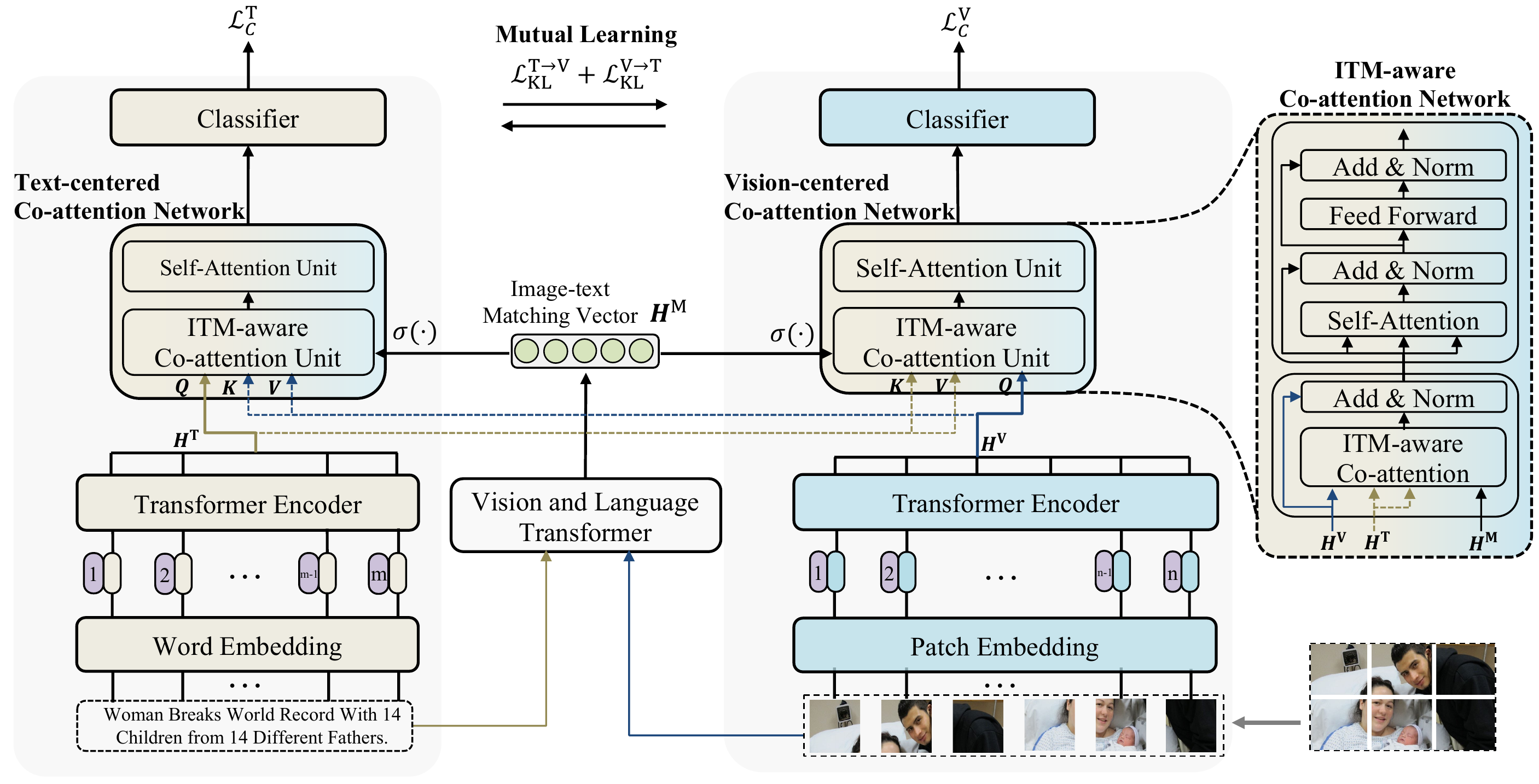}
    \caption{The architecture of our MMCAN model. The right part illustrates the ITM-aware co-attention network focused on image.}
    \label{fig:MMCAN}
    \vspace{-1em}
\end{figure*}

As shown in Figure \ref{fig:MMCAN}, the proposed model has four major modules: text and image encoder, multimodal matching-aware co-attention networks, fake news classifiers, and mutual learning. 
Given news with text and image, we first utilize two different sub-models to extract features from text and image. Then the multi-modality features are fused through two multimodal matching-aware co-attention networks respectively centered on text (text features as query) and image (image features as query). After that, the output of the co-attention networks is used for fake news classification. Mutual learning is applied between the two co-attention networks  to enable mutual knowledge distillation for collaboratively improving fake news detection.

\subsection{Text and Image Encoder}
We learn the representations of text and image with a Transformer encoder \cite{DBLP:conf/nips/VaswaniSPUJGKP17}, which is able to capture the intra-modality interactions.

\textbf{Text Encoder.}
In order to accurately model the semantic information of the text $T$ in a piece of news and avoid word ambiguity, we employ the pre-trained BERT \cite{DBLP:conf/naacl/DevlinCLT19} to obtain word embeddings.
Specifically, the text $T$ is tokenized into a sequence of $m$ word tokens and we can obtain the embeddings $\bm{E}^\text{T} \in \mathbb{R}^{m \times d_t}$ by the last BERT encoding layer, where $d_t$ is the dimension of the word embedding. 
To capture the intra-modality interactions among words, we adopt a standard Transformer encoder layer \cite{DBLP:conf/nips/VaswaniSPUJGKP17} composed of a multi-head self-attention and a Feed Forward Network (FFN) to learn the text representation $\bm{H}^\text{T}$ as follows:
\begin{equation}
    \bm{H}^\text{T}=\mbox{Transformer}(\bm{E}^\text{T} + \bm{E}^\text{T}_{pos}),
    \label{eq2}
\end{equation}
where $\bm{H}^\text{T} \in \mathbb{R}^{m \times d_t}$, and $\bm{E}^\text{T}_{pos}$ denotes the parameter-free positional embedding.


\textbf{Image Encoder.}
We also employ the Transformer, \textcolor{black}{which has proved to be effective in many visual understanding tasks \cite{DBLP:journals/tcsv/WangQCZ23,DBLP:journals/tcsv/CaoFZWY23,DBLP:journals/tcsv/DaiHZZ22}}, to extract the visual features of the news. As the standard Transformer takes a 1D sequence of token embeddings as input, we first reshape the image $\bm{V}$ into a sequence of flattened 2D patches. Then we use a trainable linear projection to flatten the patches as the input for the pre-trained ViT. In particular, we use the ViT-B/16 \cite{DBLP:conf/iclr/DosovitskiyB0WZ21} pre-trained on ImageNet to get the patch embeddings $\bm{E}^\text{V} \in \mathbb{R}^{n \times d_v}$, where $n$ is the number of patches and $d_v$ is the dimension of the patch embeddings.
Analogously, a Transformer encoder layer is adopted for internal interactions of visual modality. Formally, we can get the visual representation $\bm{H}^\text{V}$ as follows:
\begin{equation}
    \bm{H}^\text{V}=\mbox{Transformer}(\bm{E}^\text{V}+\bm{E}^{\text{V}}_{pos}),
    \label{eq4}
\end{equation}
where $\bm{H}^\text{V} \in \mathbb{R}^{n \times d_v} $, and $\bm{E}^\text{V}_{pos}$ is the positional embedding \cite{DBLP:conf/nips/VaswaniSPUJGKP17}.


\subsection{ITM-aware Co-attention Networks}

In this subsection, we describe our ITM-aware co-attention networks for learning multimodal fusion features. Specifically, the co-attention networks employ the newly designed ITM-aware co-attention mechanism which can better model the inter-modality interactions guided by the alignment of image and text.
As shown in Figure \ref{fig:MMCAN}, two ITM-aware co-attention  networks respectively focused on textual and visual information are constructed for multimodal fusion representation learning. 
Taking the ITM-aware co-attention network focused on image as an example, the detail of the co-attention network is illustrated in the right part of Figure \ref{fig:MMCAN}. It consists of  an \textit{ITM-aware co-attention unit} for inter-modality interactions and a \textit{self-attention unit} for further information interaction to detect fake news.

\begin{figure*}[!tb]
    \centering
    \includegraphics[width=0.8\textwidth]{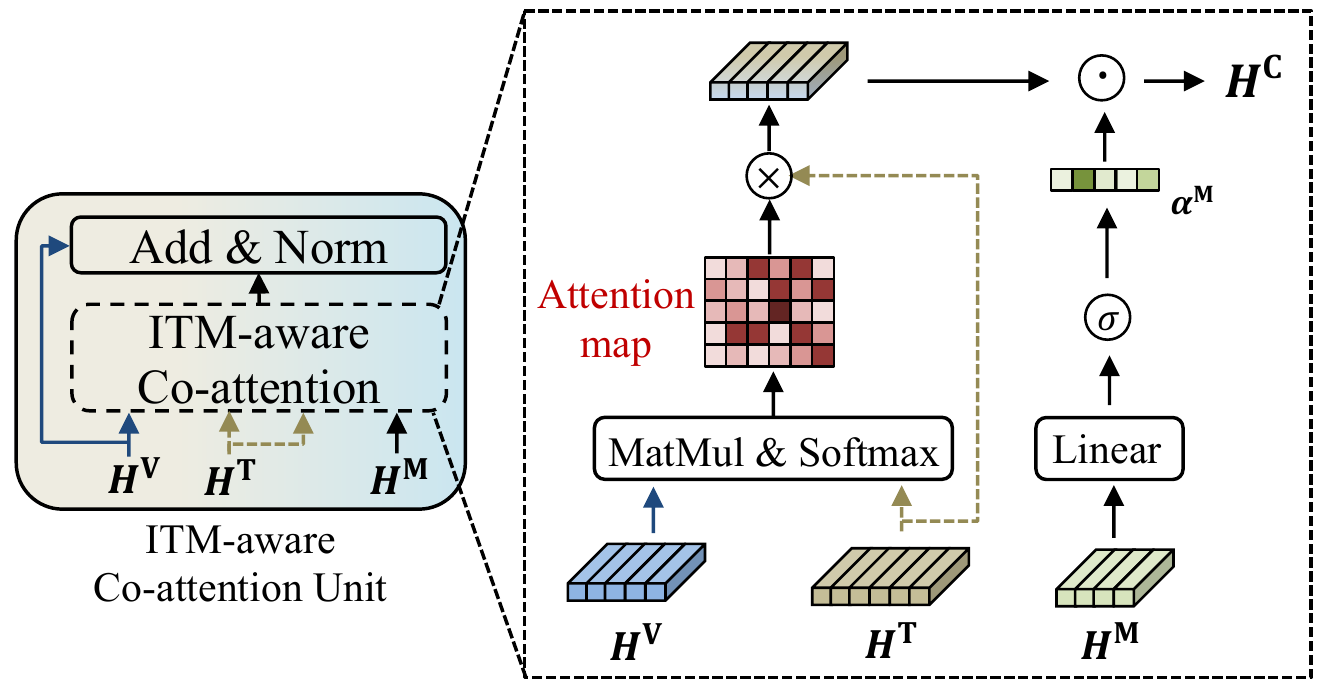}
    \caption{{The architecture of the \textcolor{black}{ITM-aware} co-attention unit. }}
    \label{fig:matching}
\end{figure*}

\textbf{ITM Representation}. To capture the alignment of image and text in our co-attention mechanism for learning better multimodal fusion features, we first leverage a vision-language pre-trained model to obtain the image-text matching representation. 
In this work, we use the pre-trained Vision and Language Transformer (ViLT)  which takes the ITM task as one of the pre-training tasks. Formally, given the original text $T$ and the attached image $V$, we obtain the ITM representation based on the ITM head of the pre-trained ViLT:
\begin{equation}
\bm{H}^\text{M}=\mbox{ITM-head}(\mbox{ViLT}(T,V)).
\end{equation}
It implies the alignment of image and text, where the ITM-head is a single linear layer projecting the pooled features to logits over binary class.
\textbf{ITM-aware Co-attention Unit}. 
In order to learn better multimodal fusion features guided by the alignment of image and text, we design a new image-text matching-aware co-attention mechanism for cross-modality interaction. Take the vision-centered co-attention network as an example, as shown in Figure \ref{fig:matching}, the queries are \textcolor{black}{from} the visual features  $\bm{H}^\text{V}$, and the keys and values are \textcolor{black}{from} the textual features $\bm{H}^\text{T}$. They are passed as inputs to the multi-head co-attention to model interactions between modalities. Formally, the inputs of the $i$-th head of co-attention are transformed as follows:
\begin{equation}
    \bm{Q}_i,\bm{K}_i,\bm{V}_i=\bm{H}^\text{V}\bm{W}_{Q_i},\bm{H}^\text{T}\bm{W}_{K_i},\bm{H}^\text{T}\bm{W}_{V_i},
    \label{eq6}
\end{equation}
where $\bm{W}_{Q_i} \in \mathbb{R}^{d_v \times d_v/h}$ ($h$ denotes the number of heads) is the projection matrix for queries, and $\{\bm{W}_{K_i},\bm{W}_{V_i} \} \in \mathbb{R}^{d_t \times d_t/h}$ are the projection matrices for keys and values, respectively. 

\textcolor{black}{Afterwards, we calculate the multi-head attention vector based on query, key, and value matrices to capture cross-modality correlations.
To guide the learning of multimodal fusion features with the image-text alignment, we employ a gating function on the ITM representation $\bm{H}^\text{M}$ to obtain a soft weight distribution $\alpha^\text{M}$, which is used to adjust the multi-head attention vector via element-wise multiplication.}
Formally, the output features $\bm{H}^\text{C}$ of the ITM-aware co-attention mechanism can be calculated as follows:
\begin{equation}
\left\{
\begin{array}{lr}
    \label{equation:attention}
    \mbox{Att}_i(\bm{H}^\text{V},\bm{H}^\text{T})=\mbox{softmax}(\frac{\bm{Q}_i\bm{K}^{\mathsf{T}}_i}{\sqrt{d_t/M}})\bm{V}_i,\vspace{1ex} \\
    \mbox{MH-Att}(\bm{H}^\text{V},\bm{H}^\text{T})=[\mbox{Att}_1 \oplus \cdots \oplus \mbox{Att}_M]\bm{W}^{'}, \vspace{1.5ex} \\
    \alpha^\text{M}=\sigma(\bm{H}^\text{M}\bm{W}^\text{M}+\bm{b}^\text{M}), \vspace{1.5ex} \\
    \bm{H}^\text{C} = \alpha^\text{M} \odot \mbox{MH-Att}(\bm{H}^\text{V},\bm{H}^\text{T}),
\end{array}
\right.
\end{equation}
where $\mbox{Att}_i$ refers to the $i$-th head of multi-head co-attention, $\bm{W}^{'} \in \mathbb{R}^{d_v \times d_v}, \bm{W}^{\text{M}} \in \mathbb{R}^{d_m \times n}$ are the weight matrices, $\bm{b}^\text{M}$ is the bias term, $\oplus$ denotes the concatenation operation, and $\odot$ denotes element-wise multiplication.

In order to address the degradation problem and normalize the distributions of intermediate layers, we wrap a residual connection and LayerNorm (LN) around the matching-aware co-attention. Formally, we can obtain the output of the ITM-aware co-attention unit:
\begin{align}
    &\widetilde{\bm{O}}^{\text{V}}=\mbox{LN}(\bm{H}^\text{V}+\bm{H}^\text{C}).
    \label{eq8}
\end{align}

\textbf{Self-Attention Unit.} After the ITM-aware co-attention, self-attention is applied to enable further interactions between modalities for the fake news classifier. Finally, we can obtain the vision-centered multimodal fusion features $\bm{O}^\text{V}$:
\begin{equation}
\left\{
\begin{array}{lr}
    \bm{H}^\text{V}_\text{S} = \mbox{LN}(\widetilde{\bm{O}}^{\text{V}} + \mbox{MH-Att}(\widetilde{\bm{O}}^{\text{V}},\widetilde{\bm{O}}^{\text{V}})), \vspace{2ex} \\
    \bm{O}^\text{V} = \mbox{LN}(\bm{H}^\text{V}_\text{S} + \mbox{FFN}(\bm{H}^\text{V}_\text{S})).
\end{array}
\right.
\label{eq9}
\end{equation}

Analogously, the text-centered ITM-aware co-attention network takes the text representation $\bm{H}^\text{T}$ as queries, and the image representation $\bm{H}^\text{V}$ as keys and values, and outputs the multimodal fusion features $\bm{O}^\text{T}$ focused on text.


\subsection{Fake News Classifier}
For the obtained multimodal fusion features, we exploit a fully connected layer followed by a softmax function to predict the authenticity of news:
\begin{gather}
    \bm{P}^{\text{T}(\text{V})}=\mbox{softmax}(\bm{W}^{\text{T}(\text{V})}\bm{O}^{\text{T}(\text{V})}+\bm{b}^{\text{T}(\text{V})}),
\end{gather}
where $\bm{P}^{\text{T}(\text{V})}$ denotes the predicted probabilities based on text(vision)-centered multimodal fusion features $\bm{O}^{\text{T}(\text{V})}$. We employ cross entropy to calculate the classification loss:
 \begin{align}
    \mathcal{L}_\text{C}^{\text{T}(\text{V})}=\sum_{i=1}^{N}&-[\bm{y}_i \mbox{log}(\bm{P}_i^{\text{T}(\text{V})}) \nonumber \\ 
    &+ (1-\bm{y}_i)\mbox{log}(1-\bm{P}_i^{\text{T}(\text{V})})],
    \label{eq11}
\end{align}
where $N$ is the number of news, $\bm{y}_i$ and $\bm{P}_i^{\text{T}(\text{V})}$ respectively denote the ground-truth label and the predicted probability of the $i$-th news based on the text(vision)-centered co-attention network.



\subsection{Mutual Learning}
As both text- and vision-centered fake news classifiers refer to the same learning objective, there should be certain intrinsic consistency between the two classifiers \cite{DBLP:conf/cvpr/ZhangXHL18,DBLP:conf/sigir/WenSYZN21}. In view of this, we adopt the mutual learning strategy to enable knowledge distillation between the two classifiers for collaboratively improving fake news detection. 

Specifically, we force the two classifiers to imitate each other in the final predicted probabilities \cite{DBLP:conf/cvpr/ZhangXHL18}  as shown in the top of Figure \ref{fig:MMCAN}.
We adopt Kullback Leibler (KL) Divergence to quantify the consistency of the two classifiers' predictions  $\bm{P}^\text{T}$ and $\bm{P}^\text{V}$:
\begin{gather}
    D_{\rm KL}(\bm{P}^{\text{T}(\text{V})}||\bm{P}^{\text{V}(\text{T})})=\sum_{i=1}^{N} \bm{P}_i^{\text{T}(\text{V})} \mbox{log}(\frac{\bm{P}_i^{\text{T}(\text{V})}}{\bm{P}_i^{\text{V}(\text{T})}}).
    \label{eq12}
\end{gather}

The mutual learning loss $\mathcal{L}^{\text{T}(\text{V}) \to \text{V}(\text{T})}_{\rm KL}$ for regularizing the vision (text) centered co-attention network to imitate the text (vision) centered co-attention network can be denoted as:
\begin{equation}
    \mathcal{L}^{\text{T}(\text{V}) \to \text{V}(\text{T})}_{\rm KL} = D_{\rm KL}(\bm{P}^{\text{V}(\text{T})}||\bm{P}^{\text{T}(\text{V})}).
    \label{eq13}
\end{equation}

The final objective function for MMCAN becomes:
\begin{equation}
    \mathcal{L}=\mathcal{L}_\text{C}^\text{T}+\mathcal{L}_\text{C}^\text{V}+\lambda_{\rm KL}(\mathcal{L}^{\text{T} \to \text{V}}_{\rm KL}+\mathcal{L}^{\text{V} \to \text{T}}_{\rm KL}) \label{eq14},
\end{equation}
where $\lambda_{\rm KL}$ is used to balance the classification and mutual learning losses.
For the inference stage, we will average the prediction probabilities of the text- and vision-centered fake news classifiers to obtain the final prediction probabilities.

\section{Experiments}
In this section, we evaluate the effectiveness of our  MMCAN. 

\subsection{Experimental Settings}
\begin{table}[!htbp]
\centering
\caption{The Statistics of Three Benchmark  Datasets.}
\scalebox{1.0}{\begin{tabular}{|c|c|c|}
\hline
Datasets & \# of Real News & \# of Fake News \\ \hline
Weibo    & 4779            & 4749            \\ \hline
Twitter  & 6026            & 7898            \\ \hline
Pheme    & 1972            & 3670            \\ \hline
\end{tabular}}
\label{tabel:tabel1}
\end{table}

\noindent\textbf{Datasets.} To evaluate the performance of the proposed MMCAN, we utilize three widely used datasets: Weibo, Twitter  and Pheme. The \textbf{Weibo} dataset (in Chinese) \cite{DBLP:conf/mm/JinCGZL17} is \textcolor{black}{collected and verified by Xinhua News Agency and Weibo,}
where each post contains text, attached image, and social information. All the fake news is crawled from May, 2012 to January, 2016 and verified by the official fake news debunking system of Weibo. The \textbf{Twitter} dataset \cite{mediaeval} is released for Verifying Multimedia Use task\footnote{http://www.multimediaeval.org/mediaeval2016/.} which aims to detect fake multimedia content on social media. Each tweet in the dataset involves the text, image and social context information. The \textbf{Pheme} dataset \cite{DBLP:conf/socinfo/ZubiagaLP17} is collected based on 5 breaking news, and each news contains a set of claims.

Table \ref{tabel:tabel1} shows the detailed statistics of the three benchmark datasets. In addition, the Weibo dataset contains 9,528 unique images, the Twitter dataset contains 514 unique images and the Pheme dataset contains 3,670 unique images. 

We adopt the data preprocessing and split the data into training and testing sets as the same with \cite{DBLP:conf/www/KhattarG0V19,DBLP:conf/sigir/QianWHFX21}. \textcolor{black}{If the data split of certain baselines is inconsistent with ours or the experimental results on certain datasets are not given in the original papers, we reproduce the corresponding experimental results according to their released codes.}

\begin{table*}[!htb]
\centering
\caption{Performance comparison among different models on three datasets. The best results are in bold. We mark the results of MMCAN with $^*$ if they exceed the results of the strong baseline MFAN with statistical significance (p$<$0.05).
$\dag$  denotes our reproduced results based on the officially released codes and the rest of the baseline results are directly taken from the papers.}
\scalebox{1.0}{
\begin{tabular}{|c|c|c|ccc|ccc|}
\hline
                           &                          &                                       & \multicolumn{3}{c|}{Fake News}                                                                                                                                  & \multicolumn{3}{c|}{Real News}                                                                                                                                  \\ \cline{4-9} 
\multirow{-2}{*}{Datasets} & \multirow{-2}{*}{Models} & \multirow{-2}{*}{Accuracy}            & \multicolumn{1}{c|}{Precision}                             & \multicolumn{1}{c|}{Recall}                                & F1                                    & \multicolumn{1}{c|}{Precision}                             & \multicolumn{1}{c|}{Recall}                                & F1                                    \\ \hline
                           & GRU                      & 0.702                                 & \multicolumn{1}{c|}{0.671}                                 & \multicolumn{1}{c|}{0.794}                                 & 0.727                                 & \multicolumn{1}{c|}{0.747}                                 & \multicolumn{1}{c|}{0.609}                                 & 0.671                                 \\
                           & {ViLT}                     & 0.832                                 & \multicolumn{1}{c|}{0.831}                                 & \multicolumn{1}{c|}{0.837}                                 & 0.834                                 & \multicolumn{1}{c|}{0.834}                                 & \multicolumn{1}{c|}{0.827}                                 & 0.830                                 \\
                           & MVAE                     & 0.824                                 & \multicolumn{1}{c|}{0.854}                                 & \multicolumn{1}{c|}{0.769}                                 & 0.809                                 & \multicolumn{1}{c|}{0.802}                                 & \multicolumn{1}{c|}{0.879}                                 & 0.870                                 \\
                           & SpotFake+                 & 0.870                                 & \multicolumn{1}{c|}{0.887}                                 & \multicolumn{1}{c|}{0.849}                                 & 0.868                                 & \multicolumn{1}{c|}{0.855}                                 & \multicolumn{1}{c|}{0.892}                                 & 0.873                                 \\
                           & SAFE                 & 0.763                                 & \multicolumn{1}{c|}{0.833}                                 & \multicolumn{1}{c|}{0.659}                                 & 0.736                                 & \multicolumn{1}{c|}{0.717}                                 & \multicolumn{1}{c|}{0.868}                                 & 0.785                                 \\
                           & CAFE                     & {0.840}                           & \multicolumn{1}{c|}{{0.855}}                                 & \multicolumn{1}{c|}{{0.830}}                           & {0.842}                           & \multicolumn{1}{c|}{{0.825}}                           & \multicolumn{1}{c|}{0.851}                                 & {0.837}                           \\ 
                           & HMCAN                    & 0.885                                 & \multicolumn{1}{c|}{{0.920}}                        & \multicolumn{1}{c|}{0.845}                                 & 0.881                                 & \multicolumn{1}{c|}{0.856}                                 & \multicolumn{1}{c|}{{0.926}}                        & 0.890                                 \\
                           & MCAN                     & {0.899}                           & \multicolumn{1}{c|}{{0.913}}                                 & \multicolumn{1}{c|}{{0.889}}                           & {0.901}                           & \multicolumn{1}{c|}{{0.884}}                           & \multicolumn{1}{c|}{0.909}                                 & {0.897}                           \\
                           & $\text{MFAN}^\dag$                     & {0.891}                           & \multicolumn{1}{c|}{{0.942}}                                 & \multicolumn{1}{c|}{{0.835}}                           & {0.885}                           & \multicolumn{1}{c|}{{0.850}}                           & \multicolumn{1}{c|}{0.948}                                 & {0.896}                           \\ \cline{2-9}
                           
                           & MMCAN-Res                & {0.906}                              & \multicolumn{1}{c|}{0.916}                                & \multicolumn{1}{c|}{{0.897}}                              & {0.906}                                 & \multicolumn{1}{c|}{{0.898}}                                 & \multicolumn{1}{c|}{{0.916}}                                 & {0.907}                                 \\
\multirow{-11}{*}{Weibo}   & MMCAN                   & ${\textbf{0.911}}^*$ & \multicolumn{1}{c|}{{{0.913}}}    & \multicolumn{1}{c|}{{0.910}} & ${\textbf{0.912}}^*$ & \multicolumn{1}{c|}{{0.909}} & \multicolumn{1}{c|}{{{0.912}}}    & ${\textbf{0.911}}^*$ \\ \hline
                           & GRU                      & 0.634                                 & \multicolumn{1}{c|}{0.581}                                 & \multicolumn{1}{c|}{0.812}                                 & 0.677                                 & \multicolumn{1}{c|}{0.758}                                 & \multicolumn{1}{c|}{0.502}                                 & 0.604                                 \\
                           & {ViLT}                     & 0.759                                 & \multicolumn{1}{c|}{0.767}                                 & \multicolumn{1}{c|}{0.393}                                 & 0.520                                 & \multicolumn{1}{c|}{0.757}                                 & \multicolumn{1}{c|}{{0.941}}                                 & 0.839                                 \\
                           & MVAE                     & 0.745                                 & \multicolumn{1}{c|}{0.801}                                 & \multicolumn{1}{c|}{0.719}                                 & 0.758                                 & \multicolumn{1}{c|}{0.689}                                 & \multicolumn{1}{c|}{0.777}                                 & 0.730                                 \\
                           & SpotFake+                 & 0.790                                 & \multicolumn{1}{c|}{0.793}                                 & \multicolumn{1}{c|}{0.827}                                 & 0.810                                 & \multicolumn{1}{c|}{0.786}                                 & \multicolumn{1}{c|}{0.747}                                 & 0.766                                 \\
                           & SAFE                 & 0.766                                 & \multicolumn{1}{c|}{0.777}                                 & \multicolumn{1}{c|}{0.794}                                 & 0.786                                 & \multicolumn{1}{c|}{0.752}                                 & \multicolumn{1}{c|}{0.731}                                 & 0.742                                 \\
                           & CAFE                     & {0.806}                           & \multicolumn{1}{c|}{{0.807}}                                 & \multicolumn{1}{c|}{{0.799}}                           & {0.803}                           & \multicolumn{1}{c|}{{0.805}}                           & \multicolumn{1}{c|}{0.813}                                 & {0.809}                           \\ 
                           & HMCAN                    & {0.897}                           & \multicolumn{1}{c|}{0.971}                        & \multicolumn{1}{c|}{0.801}                                 & {0.878}                           & \multicolumn{1}{c|}{{0.853}}                           & \multicolumn{1}{c|}{0.979}                        & {0.912}                           \\
                           & MCAN                     & 0.809                                 & \multicolumn{1}{c|}{{{0.889}}}                           & \multicolumn{1}{c|}{0.765}                                 & 0.822                                 & \multicolumn{1}{c|}{0.732}                                 & \multicolumn{1}{c|}{0.871}                                 & 0.795                                 \\
                           & $\text{MFAN}^\dag$                     & {0.925}                           & \multicolumn{1}{c|}{{0.835}}                                 & \multicolumn{1}{c|}{{{0.965}}}                           & {0.896}                           & \multicolumn{1}{c|}{{{0.981}}}                           & \multicolumn{1}{c|}{0.906}                                 & {0.942}                           \\ \cline{2-9}
                           
                           & MMCAN-Res                & {0.933}                                 & \multicolumn{1}{c|}{0.861}                                 & \multicolumn{1}{c|}{{0.950}}                                 & {0.903}                                 & \multicolumn{1}{c|}{{0.974}}                                 & \multicolumn{1}{c|}{0.924}                                 & {0.948}                                 \\
\multirow{-11}{*}{Twitter} & MMCAN                   & ${\textbf{0.943}}^*$ & \multicolumn{1}{c|}{{0.869}}          & \multicolumn{1}{c|}{{0.976}} & ${ \textbf{0.919}}^*$ & \multicolumn{1}{c|}{{0.987}} & \multicolumn{1}{c|}{{{0.927}}}    & ${\textbf{0.956}}^*$ \\ \hline
                           & GRU                      & 0.832                                 & \multicolumn{1}{c|}{0.782}                                 & \multicolumn{1}{c|}{0.712}                                 & 0.745                                 & \multicolumn{1}{c|}{0.855}                                 & \multicolumn{1}{c|}{0.896}                                 & 0.865                                 \\
                           & {ViLT}                     & 0.821                                 & \multicolumn{1}{c|}{0.659}                                 & \multicolumn{1}{c|}{0.815}                                 & 0.729                                 & \multicolumn{1}{c|}{0.914}                                 & \multicolumn{1}{c|}{0.824}                                 & 0.867                                 \\
                           & MVAE                     & 0.852                                 & \multicolumn{1}{c|}{0.806}                                 & \multicolumn{1}{c|}{0.719}                                 & 0.760                                 & \multicolumn{1}{c|}{0.871}                                 & \multicolumn{1}{c|}{{0.917}}                           & 0.893                                 \\
                           & SpotFake+                 & 0.800                                 & \multicolumn{1}{c|}{0.730}                                 & \multicolumn{1}{c|}{0.668}                                 & 0.697                                 & \multicolumn{1}{c|}{0.832}                                 & \multicolumn{1}{c|}{0.869}                                 & 0.850                                 \\
                           & SAFE                 & 0.811                                 & \multicolumn{1}{c|}{0.827}                                 & \multicolumn{1}{c|}{0.559}                                 & 0.667                                 & \multicolumn{1}{c|}{0.806}                                 & \multicolumn{1}{c|}{{0.940}}                                 & 0.866                                 \\
                           & $\text{CAFE}^\dag$                 & 0.861                                 & \multicolumn{1}{c|}{0.812}                                 & \multicolumn{1}{c|}{0.645}                                 & 0.719                                 & \multicolumn{1}{c|}{0.875}                                 & \multicolumn{1}{c|}{{0.943}}                                 & 0.908                                 \\
                           & HMCAN                    & {0.881}                           & \multicolumn{1}{c|}{{{0.830}}}                           & \multicolumn{1}{c|}{{{0.838}}}                        & \textbf{0.834}                        & \multicolumn{1}{c|}{{0.910}}                           & \multicolumn{1}{c|}{0.905}                                 & {0.907}                           \\
                           & $\text{MCAN}^\dag$                 & 0.865                                 & \multicolumn{1}{c|}{0.790}                                 & \multicolumn{1}{c|}{0.680}                                 & 0.731                                 & \multicolumn{1}{c|}{0.887}                                 & \multicolumn{1}{c|}{{0.933}}                                 & 0.910                                 \\
                           & $\text{MFAN}^\dag$                    & {0.888}                           & \multicolumn{1}{c|}{{0.771}}                                 & \multicolumn{1}{c|}{{0.846}}                           & {0.807}                           & \multicolumn{1}{c|}{{0.939}}                           & \multicolumn{1}{c|}{0.905}                                 & {0.922}                           \\ \cline{2-9}
                           & MMCAN-Res                & {0.890}                                 & \multicolumn{1}{c|}{0.803}                                 & \multicolumn{1}{c|}{{0.794}}                                 & 0.799                                 & \multicolumn{1}{c|}{{0.922}}                                 & \multicolumn{1}{c|}{{0.926}}                                 & {0.924}                                 \\
\multirow{-11}{*}{Pheme}    & MMCAN                   & ${\textbf{0.903}}^*$ & \multicolumn{1}{c|}{{0.855}} & \multicolumn{1}{c|}{{{0.777}}}    & ${ {{0.814}}}^*$    & \multicolumn{1}{c|}{{{0.918}}} & \multicolumn{1}{c|}{{0.950}} & ${\textbf{0.934}}^*$ \\ \hline
\end{tabular}
}
\label{tabel:tabel2}
\end{table*}

\noindent\textbf{Implementation Details.} 

We use the pre-trained BERT \cite{DBLP:conf/naacl/DevlinCLT19} to get the word embeddings with $768$ dimensions.
For the input image, we resize it to $224 \times 224$ and employ ViT-B/16 \cite{DBLP:conf/iclr/DosovitskiyB0WZ21} pre-trained on ImageNet to get the patch embeddings with 768 dimensions.
The number of attention heads $h$ in MMCAN is 8.
We use AdamW \cite{DBLP:conf/iclr/LoshchilovH19} optimizer with an initial learning rate of 0.001 and weight decay of 0.01 for model optimization.  The model is trained for 80 epochs with early stopping to prevent overfitting. We empirically set the batch size as 64 and the trade-off hyper-parameters  $\lambda_{\rm KL}$=$0.01$.
To mitigate overfitting, we apply dropout with the rate as 0.4. 
It should be noticed that in our model, the original parameters of ViLT are frozen, and other parameters are trainable and initialized randomly. 
For the news pieces containing multiple images (more than one image), we follow \cite{DBLP:conf/www/0003LZSLTS22} to randomly select one image.
Note that ViLT takes English text as input, thus  we pre-translate the text in the Weibo dataset into English \textcolor{black}{via the googletrans library\footnote{https://github.com/ssut/py-googletrans.} }when applying ViLT.

\noindent\textbf{{Evaluation Metrics.}}
We employ Accuracy as the evaluation metric for the fake news detection task. Considering the effect of label distribution imbalance, we also report the Precision, Recall, and F1 score of all the models for both fake news and real news following previous works \cite{DBLP:conf/www/KhattarG0V19,DBLP:conf/acl/WuZZWX21,DBLP:conf/sigir/QianWHFX21}.







\begin{table*}[!tb]
\centering
\renewcommand\arraystretch{1.0}
\caption{Performance comparison among different variants of MMCAN on Weibo, Twitter and Pheme Datasets. \textcolor{black}{The best results are in bold.}}
\scalebox{1.0}{\begin{tabular}{|c|c|c|ccc|ccc|}
\hline
\multirow{2}{*}{Datasets} & \multirow{2}{*}{Models} & \multirow{2}{*}{Accuracy} & \multicolumn{3}{c|}{Fake news}                                       & \multicolumn{3}{c|}{Real news}                                       \\ \cline{4-9} 
                          &                         &                           & \multicolumn{1}{c|}{Precision} & \multicolumn{1}{c|}{Recall} & F1    & \multicolumn{1}{c|}{Precision} & \multicolumn{1}{c|}{Recall} & F1    \\ \hline
\multirow{6}{*}{Weibo}  
& MMCAN-w/o-Match                & 0.901                     & \multicolumn{1}{c|}{0.907}     & \multicolumn{1}{c|}{0.895}  & 0.901 & \multicolumn{1}{c|}{0.895}     & \multicolumn{1}{c|}{0.907}  & 0.901 \\
& MMCAN-w/-T                & 0.900                     & \multicolumn{1}{c|}{0.910}     & \multicolumn{1}{c|}{0.891}  & 0.900 & \multicolumn{1}{c|}{0.891}     & \multicolumn{1}{c|}{0.910}  & 0.901 \\
                          & MMCAN-w/-V                & 0.897                     & \multicolumn{1}{c|}{0.897}     & \multicolumn{1}{c|}{0.899}  & 0.898 & \multicolumn{1}{c|}{0.897}     & \multicolumn{1}{c|}{0.895}  & 0.896 \\
                          & MMCAN-Concat              & 0.908                     & \multicolumn{1}{c|}{0.907}     & \multicolumn{1}{c|}{0.911}  & 0.909 & \multicolumn{1}{c|}{0.909}     & \multicolumn{1}{c|}{0.906}  & 0.907 \\
                          & MMCAN-Avg              & 0.902                     & \multicolumn{1}{c|}{0.934}     & \multicolumn{1}{c|}{0.866}  & 0.899 & \multicolumn{1}{c|}{0.874}     & \multicolumn{1}{c|}{0.938}  & 0.905 \\
                          
                          & MMCAN                  & \textbf{0.911}                     & \multicolumn{1}{c|}{0.913}     & \multicolumn{1}{c|}{0.910}  & \textbf{0.912} & \multicolumn{1}{c|}{0.909}     & \multicolumn{1}{c|}{0.912}  & \textbf{0.911} \\ \hline
\multirow{6}{*}{Twitter} 
& MMCAN-w/o-Match                & 0.921                         & \multicolumn{1}{c|}{0.838}         & \multicolumn{1}{c|}{0.946}      & 0.888     & \multicolumn{1}{c|}{0.971}         & \multicolumn{1}{c|}{0.909}      & 0.939     \\
& MMCAN-w/-T                & 0.913                     & \multicolumn{1}{c|}{0.813}     & \multicolumn{1}{c|}{0.958}  & 0.880 & \multicolumn{1}{c|}{0.977}     & \multicolumn{1}{c|}{0.891}  & 0.932 \\
                          & MMCAN-w/-V                & 0.919                     & \multicolumn{1}{c|}{0.841}     & \multicolumn{1}{c|}{0.929}  & 0.883 & \multicolumn{1}{c|}{0.963}     & \multicolumn{1}{c|}{0.913}  & 0.937 \\
                          & MMCAN-Concat                & 0.931                     & \multicolumn{1}{c|}{0.838}     & \multicolumn{1}{c|}{0.980}  & 0.903 & \multicolumn{1}{c|}{0.989}     & \multicolumn{1}{c|}{0.906}  & 0.946 \\
                          & MMCAN-Avg              & 0.929                     & \multicolumn{1}{c|}{0.852}     & \multicolumn{1}{c|}{0.952}  & 0.899 & \multicolumn{1}{c|}{0.975}     & \multicolumn{1}{c|}{0.918}  & 0.945 \\
                          
                          & MMCAN                  & \textbf{0.943}                     & \multicolumn{1}{c|}{0.869}     & \multicolumn{1}{c|}{0.976}  & \textbf{0.919} & \multicolumn{1}{c|}{0.987}     & \multicolumn{1}{c|}{0.927}  & \textbf{0.956} \\ \hline
\multirow{6}{*}{Pheme}  
& MMCAN-w/o-Match                & 0.893                     & \multicolumn{1}{c|}{0.859}     & \multicolumn{1}{c|}{0.731}  & 0.790 & \multicolumn{1}{c|}{0.903}     & \multicolumn{1}{c|}{0.954}  & 0.928 \\
& MMCAN-w/-T                & 0.884                     & \multicolumn{1}{c|}{0.792}     & \multicolumn{1}{c|}{0.783}  & 0.787 & \multicolumn{1}{c|}{0.918}     & \multicolumn{1}{c|}{0.922}  & 0.920 \\
                          & MMCAN-w/-V                & 0.881                     & \multicolumn{1}{c|}{0.846}     & \multicolumn{1}{c|}{0.691}  & 0.761 & \multicolumn{1}{c|}{0.890}     & \multicolumn{1}{c|}{0.952}  & 0.920 \\
                          & MMCAN-Concat                & 0.896                     & \multicolumn{1}{c|}{0.843}     & \multicolumn{1}{c|}{0.766}  & 0.802 & \multicolumn{1}{c|}{0.914}     & \multicolumn{1}{c|}{0.946}  & 0.930 \\
                          & MMCAN-Avg              & 0.888                     & \multicolumn{1}{c|}{0.851}     & \multicolumn{1}{c|}{0.720}  & 0.780 & \multicolumn{1}{c|}{0.900}     & \multicolumn{1}{c|}{0.952}  & 0.925 \\
                          
                          & MMCAN                  & \textbf{0.903}                     & \multicolumn{1}{c|}{0.855}     & \multicolumn{1}{c|}{0.777}  & \textbf{0.814} & \multicolumn{1}{c|}{0.918}     & \multicolumn{1}{c|}{0.950}  & \textbf{0.934} \\ \hline
\end{tabular}}
\label{tabel:tabel3}
\end{table*}

\subsection{Baselines}
We compare our MMCAN model with both unimodal and multimodal content based models as follows.

\noindent\textbf{Unimodal Models.}
We compare MMCAN with \textbf{{GRU}} \cite{DBLP:conf/ijcai/MaGMKJWC16}, which exploits the multilayer GRU network to encode the textual information for fake news detection.


\noindent\textbf{Multimodal Models.}
We compare our MMCAN with the following multimodal content based methods: 
\begin{itemize}
    \item [1)] \textbf{{ViLT}} \cite{DBLP:conf/icml/KimSK21}, \textcolor{black}{which is a multimodal pre-trained model. As a baseline method for fake news detection, we add a classification head on ViLT and fine-tune the head.}
    \item[2)] \textbf{{MVAE}} \cite{DBLP:conf/www/KhattarG0V19}, which employs a variational autoencoder coupled with a binary classifier for fake news detection. 
    \item[3)] \textbf{{SpotFake+}} \cite{DBLP:conf/aaai/SinghalKSS0K20}, which uses the pre-trained XLNet \cite{DBLP:conf/nips/YangDYCSL19} and VGG-19 to learn textual and visual features, and concatenates them for fake news detection.
    \item[4)] \textbf{{SAFE}} \cite{DBLP:conf/pakdd/ZhouWZ20}, which jointly exploits the multimodal features and cross-modal similarity of news content for fake news detection.
    \item[5)] \textbf{{CAFE}} \cite{DBLP:conf/www/0003LZSLTS22}, which learns the cross-modal ambiguity and uses it to adaptively aggregate multimodal features and unimodal features for fake news detection.
    \item[6)] \textbf{{HMCAN}} \cite{DBLP:conf/sigir/QianWHFX21}, which employs a hierarchical multimodal contextual attention network for fake news detection by jointly modeling the multimodal context information and the hierarchical semantics of text.
    \item[7)] \textbf{{MCAN}} \cite{DBLP:conf/acl/WuZZWX21}, which employs multiple co-attention layers to fuse the multimodal features for fake news detection.
    \item[8)] \textbf{{MFAN}} \cite{DBLP:conf/ijcai/ZhengZGWZ022}, which integrates textual, visual, and social network features through co-attention mechanism for fake news detection.
    \item[9)] \textbf{{MMCAN-Res}}, our MMCAN replacing the ViT encoder with the same ResNet-50 \cite{DBLP:conf/cvpr/HeZRS16} as HMCAN. 
\end{itemize}

\subsection{Results and Analysis}
\textcolor{black}{We run MMCAN with 5 random seeds and report the average performance in Table \ref{tabel:tabel2}.}
From the table, we can obtain the following observations: 
1) The proposed MMCAN outperforms all the baseline models on all the datasets in terms of accuracy and  most other metrics. Compared to the best baseline model, MMCAN improves the accuracy by about 1.2\%, 1.8\% and 1.5\% on Weibo, Twitter and Pheme datasets, respectively. 
2) Models that consider multimodal information outperform unimodal models, which confirms the advantage of integrating multiple modality information in fake news detection. 
3) Compared with the models which concatenate  multimodal features or leverage auxiliary tasks (i.e., SpotFake+ and MVAE) and the multimodal pre-trained model ViLT, models based on co-attention mechanism (i.e., HMCAN and MCAN) perform better, indicating that the co-attention mechanism can better fuse multimodal information.  
4) Our MMCAN further outperforms the co-attention based methods (i.e., HMCAN, MCAN and MFAN). Compared with methods that consider the consistency of textual and visual content (i.e., SAFE and CAFE), our MMCAN achieves substantial improvements.
We believe that  our proposed multimodal matching-aware co-attention networks with mutual knowledge distillation can learn better multimodal fusion features guided by the image-text alignment and enable  the two co-attention networks respectively centered on text and image to learn from each other for collaboratively improving fake news detection.
5) Our model variant MMCAN-Res which replaces the {ViT} encoder with ResNet still outperforms all the other baseline models in terms of accuracy and most other metrics. This demonstrates our model indeed benefits from the ITM-aware co-attention networks with mutual learning.

\begin{figure*}[!t]
    \setlength{\belowcaptionskip}{-0.4cm}
    \centering
    \includegraphics[width=0.8\textwidth]{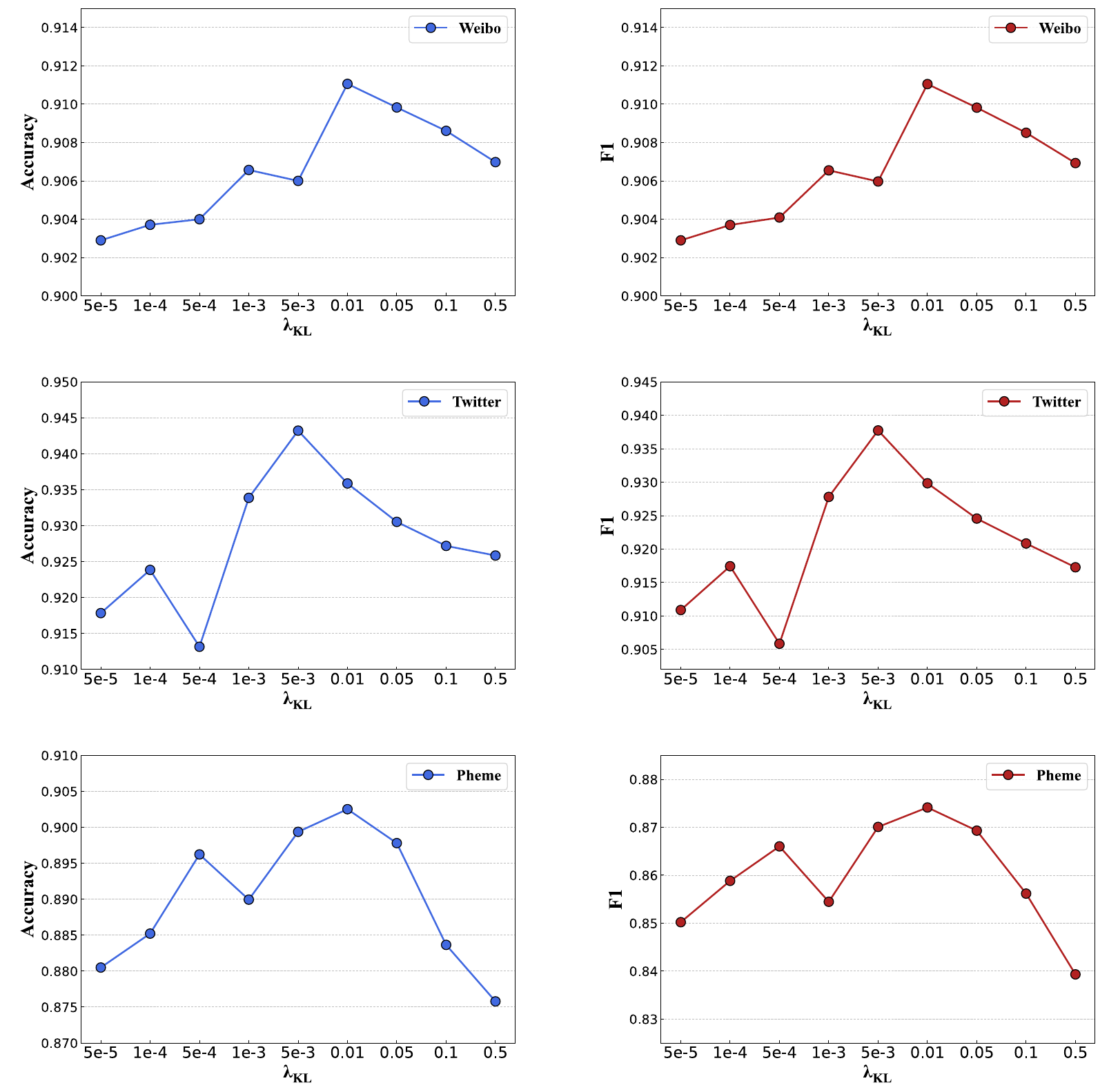}
    \caption{Impact of the value of $\lambda_{\rm KL}$ on three datasets.}
    \label{fig:KL}
\end{figure*}

\subsection{Ablation Study}
To verify the importance of each module in our MMCAN, we compare MMCAN with the following variants: 
 \begin{itemize}
\item[1)] \textbf{MMCAN-w/o-Match}, a variant of MMCAN that replaces the ITM-aware co-attention mechanism with the traditional co-attention mechanism.
\item[2)] \textbf{MMCAN-w/-T}, a variant of MMCAN based on only the co-attention network centered on text. 
\item[3)] \textbf{MMCAN-w/-V}, a variant of MMCAN based on only the co-attention network centered on image. 
\item[4)] \textbf{MMCAN-Concat}, a variant of MMCAN that concatenates the output features of MMCAN-w/-T and MMCAN-w/-V for classification.
\item[5)] \textbf{MMCAN-Avg}, a variant of MMCAN that ensembles MMCAN-w/-T and MMCAN-w/-V by averaging their prediction probabilities instead of using the mutual learning mechanism.
\end{itemize}


Table \ref{tabel:tabel3} shows the results of ablation studies. We can obtain the following observations: 
1) MMCAN consistently outperforms MMCAN-w/o-Match on three datasets, which proves the necessity of considering the image-text alignment in the co-attention mechanism for learning multimodal fusion features. 
2) Both MMCAN-w/-T and MMCAN-w/-V achieve low performance, demonstrating that individual text- or vision-centered co-attention network is suboptimal for fake news detection. On Weibo dataset, MMCAN-w/-T performs much better than MMCAN-w/-V. The reason could be that the text on Weibo is relatively long, containing more information for fake news detection \cite{DBLP:conf/acl/WuZZWX21}. 
3) MMCAN-Concat, which concatenates the text- and vision-centered multimodal features for classification, outperforms MMCAN-w/-T and MMCAN-w/-V with only text- or vision-centered multimodal features. Nevertheless, MMCAN-Concat achieves worse performance than MMCAN, demonstrating that the mutual learning enabling knowledge distillation between the text- and vision-centered co-attention networks improves the fake news detection.
4) Compared to the ensemble method MMCAN-Avg which averages the prediction probabilities of the two co-attention networks that are respectively centered on image and text, MMCAN enabling mutual knowledge distillation between them achieves significant improvements.

\subsection{Impact of the Value of $\lambda_{\rm KL}$}

To explore the impact of the balance factor $\lambda_{\rm KL}$  on the model performance, we vary $\lambda_{\rm KL}$ from $5\text{e-}5$ to $0.5$, and report the accuracy and average F1 score (the average F1 score of fake news and real news) on the three datasets in Figure \ref{fig:KL}. 
We observe that the accuracy and average F1 score of MMCAN on three datasets generally first grow as the $\lambda_{\rm KL}$ increases, and then begin to drop after  $\lambda_{\rm KL}$  is larger than a certain value. MMCAN achieves the highest value at $\lambda_{\rm KL}$=$0.005$ on Twitter dataset, and it performs best when $\lambda_{\rm KL}$=$0.01$ on both Weibo and Pheme datasets. \textcolor{black}{ The best values of $\lambda_{\rm KL}$  are relatively small. We think the reason could be that $\lambda_{\rm KL}$ works like regularization coefficient to penalize the inconsistency of the predictions of  two co-attention networks.}

\begin{figure*}[!tb]
    \centering
    \includegraphics[width=0.95\textwidth]{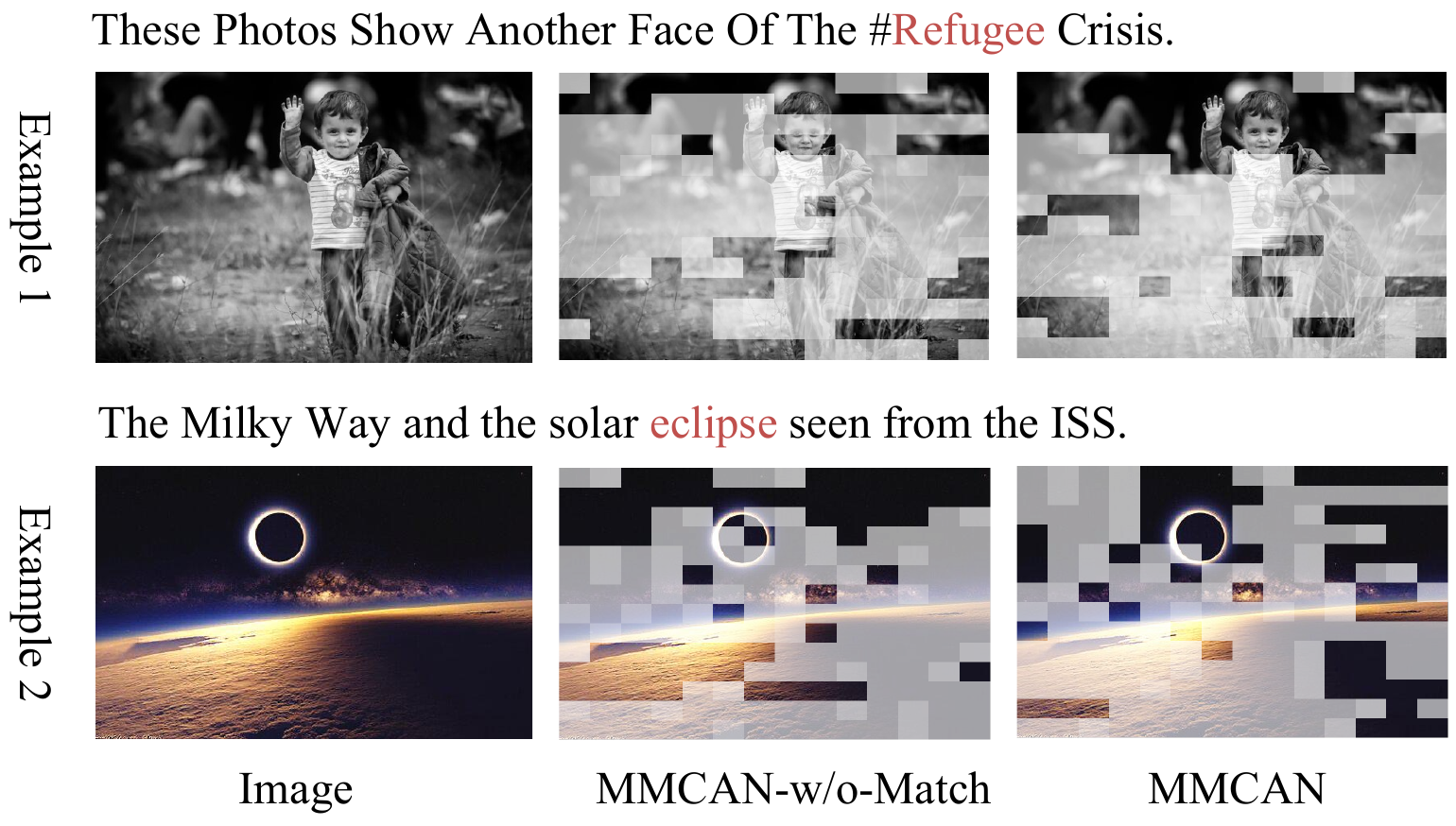}
    \caption{\textcolor{black}{Visualization of word (in red) attention weights over patches of two testing examples from Twitter dataset.
    }}
    \vspace{-1em}
    \label{fig:visualization}
\end{figure*}

\subsection{Case Study}
To gain an intuitive understanding of the ITM-aware co-attention mechanism, we visualize the word attention weights over the image patches calculated by Equation \eqref{equation:attention} in the text-centered co-attention network. For ease of illustration, we reflect the attention weights on the opacity of the patches. If the attention value is larger than the median attention weight, the opacity is set as 255; otherwise, it is set as 76. We visualize the results of  MMCAN-w/o-Match and MMCAN in Figure \ref{fig:visualization}.
As can be seen from Example 1 in Figure \ref{fig:visualization}, MMCAN is focusing on the facial area of the child as the corresponding object, which is consistent with the word `Refugee', while it is scattered all over the image in MMCAN-w/o-Match. In Example 2, we find that our MMCAN with the ITM-aware co-attention mechanism better aligns the word `eclipse' with the luminous, circular object in the image. These cases demonstrate that considering image-text alignment in co-attention mechanism can better capture the inter-modality correlations.


\section{Conclusion}

In this paper, we present novel multimodal matching-aware co-attention networks  with mutual knowledge distillation for fake news detection. The ITM-aware co-attention mechanism in MMCAN can learn better multimodal fusion features guided by the image-text alignment. 
In addition, MMCAN enables mutual knowledge distillation between co-attention networks respectively focused on text and image for collaboratively improving the fake news detection. Extensive experiments on three public benchmark datasets demonstrate the effectiveness of MMCAN. 
In future, we plan to utilize external knowledge for improving fake news detection and explore the mutual learning between different views including the knowledge-guided view.

\bibliographystyle{IEEEtran}
\bibliography{cas-refs.bib}

\end{document}